\documentclass{aa}  

\usepackage{graphicx}
\usepackage{txfonts}

\usepackage{amsmath}
\usepackage{amssymb}
\usepackage{graphicx}

\usepackage{flushend} 

\usepackage{natbib}
\newcommand{\dd}{\,\mathrm{d}}

\newcommand{\myfn}[1]{${}^{(\rm #1)}$}

\newcommand{\Mpch}{\,{\rm Mpc}\,\ifmmode h^{-1}\else $h^{-1}$\fi}
\newcommand{\Msh}{\,\ifmmode M_\odot\,h^{-1}\else $M_\odot\,h^{-1}$\fi}

\begin{document}

\title{Universal predictions of screened modified gravity\\on cluster scales}

\author{M. Gronke
          \inst{1}
          \and
          D. F. Mota\inst{1}
          \and
          H. A. Winther\inst{2}}

   \institute{Institute of Theoretical Astrophysics, University of Oslo, Postboks 1029, 0315 Oslo, Norway\\
              \email{maxbg@astro.uio.no}
         \and
            Astrophysics, University of Oxford, DWB, Keble Road, Oxford, OX1 3RH, UK\\
             }

   \date{Matches version to be published in \aap\ (Received: 27 May 2015 / Accepted: 31 August 2015)}

  \abstract
{Modified gravity models require a screening mechanism to be able to evade the stringent constraints from local gravity experiments and, at the same time, give rise to observable astrophysical and cosmological signatures. Such screened modified gravity models necessarily have dynamics determined by complex nonlinear equations that usually need to be solved on a model-by-model basis to produce predictions. This makes testing them a cumbersome process. In this paper, we investigate whether there is a common signature for all the different models that is suitable to testing them on cluster scales.
To do this we propose an observable related to the fifth force, which can be observationally related to the ratio of dynamical-to-lensing mass of a halo, and then show that the predictions for this observable can be rescaled to a near universal form for a large class of modified gravity models.
We demonstrate this using the Hu-Sawicki $f(R)$, the Symmetron, the nDGP, and the Dilaton models, as well as unifying parametrizations.
The universal form is determined by only three quantities: a strength, a mass, and a width parameter. We also show how these parameters can be derived from a specific theory. This self-similarity in the predictions can hopefully be used to search for signatures of modified gravity on cluster scales in a model-independent way.
}


   \maketitle


\section{Introduction}
A possible solution for the observed acceleration of our Universe is that general relativity (GR) is not the correct theory to describe gravity on large scales. Over the past decade, a large number of alternative gravity theories have been proposed \citep{2012PhR...513....1C}. Many of these models cannot alleviate the existing, or they introduce new problems; nevertheless, the idea that cosmic acceleration has a gravitational origin persists. 

High precision tests of gravity indicate that general relativity is the correct theory for describing gravitational physics on Earth and in the solar system \citep{Berotti2003Natur.425..374B, Will2006LRR.....9....3W, Williams2004PhRvL..93z1101W}. This places strong constraints on any theory that seeks to modify general relativity. Thus, if gravity is modified on large scales -- giving rise to cosmic acceleration -- then a way of suppressing the modifications of gravity in the well-tested regimes is required \citep{2013arXiv1312.2006K}. A key feature of screening mechanisms  currently being considered is that they are fundamentally nonlinear, thereby making the study of their cosmological effects challenging \citep{1972PhLB...39..393V,2004PhRvD..69d4026K,2010PhRvL.104w1301H}.

As previously argued in \citet{Gronke2014_dl, Gronke2014b_dl}, deviations from GR occurring in screened-modified gravity theories (SMGs) can generally be grouped into three categories: no deviations (\textit{fully screened regime}), maximum deviations (\textit{unscreened regime}), and the intermediate deviations (\textit{partially screened regime}). As a natural continuation of this idea, the main questions we would like to address in this paper follow:
\begin{itemize}
\item How do the extents and occurrences of the three regimes vary with the screened modified gravity model and its parameters?
\item Is it possible to remap observables within one model and/or between different models; i.e., are SMGs self-similar in some way?
\end{itemize}
Previous work has been carried out in this direction. \citet{2012PhLB..715...38B,2012PhRvD..86d4015B} have developed a theoretical framework in which it is possible to describe certain SMGs with two free functions. Using this framework \cite{2014arXiv1403.6492W} propose an approximate scheme to perform fast {\it N}-body simulations of SMGs by combining linear theory with a screening factor obtained from studying spherical symmetric systems. Also, \citet{Mead2014} have recently published a technique for remapping $\Lambda$CDM power spectra to the Chameleon $f(R)$ model with $\sim 3\%$ accuracy.
Complementary to that, we want to base this work on an observationally focused approach. Specifically, in this paper we focus on the behavior of screened modified gravity theories inside and in the vicinity of clusters of galaxies. In general we expect the strongest signal of modified gravity to be found in cosmological voids, but clusters are easier to work with theoretically and observationally, and they remain an important observational probe of gravity on intermediate scales \citep{Baker2015ApJ...802...63B}.

To answer the questions above, we have structured this paper as follows. In Sec.~\ref{sec:methods}, we give a general introduction to the topography of SMGs. We also picked three models as examples, which we describe in more detail. This includes the respective solving method for halo density profiles. In Sec.~\ref{sec:results}, we present the numerical results of these three models. In addition, we study the effects of varying the model parameters on the observables and try to empirically find a rescaling that counterbalances this variation. Also in this section, we develop a general analytic rescaling method and  compare it to our numerical results. In Sec.~\ref{sec:discussion}, we discuss our results and possible observational implications. We conclude in Sec.~\ref{sec:conclusion}.

Throughout this work we use $(\Omega_{m0},\,H_0)=(0.3,\,0.7)$, $M_{\rm Pl}^{-2} \equiv 8 \pi G$, $\rho_c = 3 H^2 M_{\rm Pl}^2$, $\overline{\rho_m} = \Omega_m \rho_c$ and denote values today with a subscript zero.

\section{Methods}
\label{sec:methods}

In this section we give a brief introduction to screened modified gravity theories. We present a few example models in more detail and then give a general framework for considering a large class of models. In the end we discuss the numerical implementation for solving for the field profiles of the models for the case of a Navarro-Frenk-White (NFW) density profile.

\subsection{Modified gravity theories}
\label{sec:smg-topography}

When we speak about modification of gravity or modified gravity theories, we generally mean the addition of terms to the Einstein-Hilbert action. Specifically, in scalar-tensor theories one (or multiple) additional scalar field is added with
their respective Lagrangian $L_\varphi$, and the general action reads as
\begin{align}
S =& \int\dd^4x \sqrt{-g}\left(\frac{M_{\mathrm{Pl}}^2}{2}R + L_\varphi(\varphi, \partial\varphi,\partial^2\varphi) \right)+\mathcal{S}_m(\psi^{(i)},\tilde g_{\mu\nu}),
\label{eq:action}
\end{align}
where the matter fields $\psi^{(i)}$ couple to the metric $\tilde g_{\mu\nu}\equiv A^2g_{\mu\nu}$, and $R$ is the Ricci scalar. We limit ourselves to one scalar field and a conformal coupling to matter\footnote{Disformal coupling allows for additional screening mechanisms \citep{Koivisto2012PhRvL.109x1102K}.}. Since the scalar field is coupled to matter, i.e., $A = A(\varphi)$, an extra \textit{fifth force}  arises that is in the non-relativistic limit, and per unit mass is given by 
\begin{equation}
{\bf F}_\varphi = -\vec{\nabla}\log A\;.
\end{equation}
The predictions of general relativity have been confirmed to high accuracy in the laboratory and in the solar system, which severely constrains any new such force in our (Earth's) local environment \citep{Berotti2003Natur.425..374B, Will2006LRR.....9....3W, Williams2004PhRvL..93z1101W}. If the fifth force is weak close to earth, then one would naively expect it to also be weak on cosmological scales. However, in the past decade several theories have been proposed where one is able to dynamically suppress the effect of such a fifth force in high density environments (relative to the cosmic mean) and still have interesting cosmological signatures. If this is the case, then a theory is said to possess  a \textit{\emph{screening mechanism}} \citep[for reviews see, e.g.,][]{Clifton2012,Khoury2010,Joyce2014}. 

The action Eq.~\eqref{eq:action} allows a loose mathematical classification of possible screening mechanisms:
\begin{enumerate}
\item Screening due to the scalar field value $\varphi$. If $L_\varphi$ possesses a potential $V(\varphi)$, it is possible to suppress the fifth force thanks to the value of the scalar field moving inside the potential. Physically, this can occur if the coupling strength depends on the local environment as is the case in the \textit{Symmetron} mechanism \citep{2010PhRvL.104w1301H,Hinterbichlera}. Another possibility is that the range of the fifth force depends on the environment, which is often called the \textit{Chameleon} mechanism \citep{Khourya,2004PhRvD..69d4026K,2007PhRvD..75f3501M}. A frequently considered SMG possessing the Chameleon screening is the \citet{Hu2007} $f(R)$ formulation. 
\item Screening due to derivatives of the scalar field $\partial\varphi$ and/or $\partial^2\varphi$. In this case, the screening is caused by the relation between the kinetic term of $L_\varphi$, i.e., $-\frac{1}{2}(\partial\varphi)^2$, and the rest of the scalar field Lagrangian. A popular mechanisms incorporating this idea is the \textit{\emph{Vainshtein}} mechanism \citep{1972PhLB...39..393V}, where screening depends on the value of $\partial^2\varphi$, and the k-mouflage mechanism \citep{2009IJMPD..18.2147B,2014PhRvD..90b3507B}, where the magnitude of $(\partial\varphi)^2$ decides whether a region is screened or not. Particular SMGs employing the Vainshtein Mechanism are massive gravity \citep{2014LRR....17....7D}, Galileons \citep{2009PhRvD..79f4036N}, and the DGP model \citep{2000PhLB..485..208D}. 
\end{enumerate}
A selective overview over some models and their employed screening mechanism is given in Table \ref{tab:models}.

Over the past decade there have been several studies of modified gravity on nonlinear scales, often using {\it N}-body simulations. For theories that screen via the first method described above (chameleon-like screening), there have been studies of $f(R)$ gravity \citep{Li2012JCAP...01..051L,Llinares2014a,Puchwein2013MNRAS.436..348P,2011PhRvD..83d4007Z,2008PhRvD..78l3523O,2014PhRvD..90j3505H,2012PhRvD..85l4054L}, other chameleon models \citep{2013JCAP...04..029B}, the Symmetron \citep{2012ApJ...748...61D,Llinares2014a,2012JCAP...10..002B}, and the environment-dependent dilation model \citep{2011PhRvD..83j4026B,2012JCAP...10..002B}. For models that employ the second screening method (Vainshtein-like screening), there have been studies of the DGP model \citep{Schmidt2009PhRvD..80h3505S,Schmidt2009PhRvD..80l3003S,2009PhRvD..80j4005C,2009PhRvD..80f4023K,2013JCAP...05..023L,2015arXiv150306673F,2014JCAP...07..058F}, Galileons \citep{2013JCAP...10..027B,2013JCAP...11..012L}, and k-moflage \citep{2014PhRvD..90l3521B}. This is by no means a complete list.

In spite of the differences between the different screening mechanisms and the theories employing them, some common quantities can be defined. In particular, we focus on the enhancement of the gravitational force 
\begin{equation}
\gamma \equiv G_{\rm eff} / G - 1
\end{equation}
with its theoretical maximum value $\gamma_{\rm max}$, given a specific modified gravity model.
As described in \S\ref{sec:caveats}, $\gamma$ can be related to the ratio of the dynamical to the lensing mass.

\begin{table}
\caption{Overview of screened modified gravity models}
\label{tab:models}
\begin{tabular}{lll}
\hline
\textbf{Model} & \textbf{Screening type} & \textbf{{\it N}-body code}\myfn{A}\\
\hline
$f(R)$ HS\myfn{a} & (1) Chameleon & (E), (I), (MGG), (O) \\
Dilaton &  (1) Chameleon-like & (E)\\
Symmetron & (1) Chameleon-like & (E), (I) \\
DGP & (2) Vainshtein & (E), (S), (K) \\
Galileon & (2) Vainshtein & (E), (I) \\
k-mouflage & (2) Vainshtein-like & --- \\ 
\hline
\end{tabular}
\\
\myfn{a} \citet{Hu2007} $f(R)$\\
\myfn{A}(E) \texttt{ECOSMOG} \citep{Li2012JCAP...01..051L}, (I) \texttt{ISIS} \citep{Llinares2014a}, (K) \citet{2009PhRvD..80f4023K}, (MGG) \texttt{MG-GADGET} \citep{Puchwein2013MNRAS.436..348P}, (O) \citet{2008PhRvD..78l3523O}, (S) \citet{Schmidt2009PhRvD..80h3505S,Schmidt2009PhRvD..80l3003S}
\end{table}

\subsection{Example models}
\label{sec:models}
The screened modified gravity models that we considered possess three intrinsically different screening mechanisms: the Vainshtein \citep{1972PhLB...39..393V} screening where the derivatives of the scalar field play a major role; the Symmetron \citep{Hinterbichlera,2010PhRvL.104w1301H} screening, where the coupling strength is environment dependent and goes to zero in high density regions; and the Chameleon \citep{2004PhRvD..69d4026K,Khourya} screening, where the range of the fifth force approaches zero for regions with high matter density. As an example of the latter, we focus on the \citet{Hu2007} $f(R)$ model, which incorporates the Chameleon screening mechanism.

In the following, we do not present the example models in great detail. Instead, we refer the reader to our previous work \citep{Gronke2014_dl,Gronke2014b_dl}, where we stated the full equations for the Symmetron and the \citet{Hu2007} $f(R)$ model or -- even better -- to excellent reviews, such as those by \citet{Clifton2012}, \citet{2013arXiv1312.2006K}, \citet{Joyce2014}, and, \citet{2015arXiv150404623K}.

\subsubsection{Symmetron model}
\label{sec:symmetron-model}
In the Symmetron model \citep{2010PhRvL.104w1301H,Hinterbichlera}, both the coupling function and the potential are symmetric around $\varphi = 0$  are given by
\begin{align}
A(\varphi) &= 1 + \frac{\varphi^2}{M^2} \\
V(\varphi) &= -\frac{1}{2}\mu^2\varphi^2+\frac{1}{4}\lambda\varphi^4.
\end{align}
Here, the prefactors $M$, $\mu$ and $\lambda$ can be rewritten as
\begin{align}
  M^2 &= \frac{2 \Omega_{m0} \rho_{c0} L^2}{a_{\rm ssb}^3}\\
  \mu &= \frac{1}{\sqrt{2} L}\\
  \lambda^{1/2} &=\frac{a_{\rm ssb}^3 M_{\rm Pl}}{\sqrt{8} L^3 \Omega_{m0}\rho_{c0}\beta,}
\end{align}
which leaves the Symmetron parameters to be the scale factor at symmetry breaking $a_{\rm ssb}$, the range of the fifth force in vacuum $L$, and the fiducial coupling $\beta$.

In a Friedmann-Lema\^itre-Robertson-Walter (FLRW) background and in the quasi-static limit \citep{2014PhRvD..89b3521N}, the field equation becomes
\begin{equation}\label{eq:symm_full}
\nabla^2\tilde{\varphi} = \frac{a^2}{2\lambda_0^2}\left(\tilde{\varphi}\frac{a_{\rm ssb}^3}{a^3}\frac{\rho_m}{\overline{\rho_m}} - \tilde{\varphi} + \tilde{\varphi}^3 \right)
\end{equation}
where $\tilde{\varphi} = \sqrt{\lambda}\varphi / \mu$.
The fifth force is given by
\begin{align}
F_\varphi = \frac{\varphi \nabla \varphi}{M^2.}
\label{eq:fifth_force_symm}
\end{align}
In regions of high ambient matter density ($\rho_m \gg \mu^2M^2 = \rho_{c0}/a_{\rm ssb}^3$), the field will reside close to $\phi = 0$, and the fifth force will be screened. In low-density regions, on the other hand, the fifth force can be in full operation. In the Symmetron model, the maximum enhancement of the gravitational force, $\gamma_{\rm max}$, is given by
\begin{equation}
\gamma_{\rm max} = 2 \beta^2 \left[ 1 - \left(\frac{a_{\rm ssb}}{a}\right)^3\right]\;.
\label{eq:gammamax_symm}
\end{equation}
Since $\beta$ is a free parameter, the model can, in principle, give rise to an arbitrarily large (or small) deviation. How the screening mechanism works is explained in more detail in Sec.~\ref{sect:unified}.

\subsubsection{Hu-Sawicki $f(R)$ model}
\label{sec:chameleon-fr-model}
The $f(R)$ gravity models are a group of modified gravity theories where the Ricci scalar $R$ in the Einstein-Hilbert action is replaced by a generic function $R + f(R)$. By applying a conformal transformation, the $f(R)$ action can be brought in the form of Eq.~\eqref{eq:action} \citep[see, e.g.,][]{Clifton2012,2008PhRvD..78j4021B}. This transformation\footnote{$\tilde g_{\mu\nu}=A^2(\varphi)g_{\mu\nu}$ where $A(\varphi) = \mathrm{e}^{\beta\varphi/M_{\rm Pl}}$ with $\beta = 1/\sqrt{6}$} translates $f(R)$-gravity into a scalar tensor theory. The theory screens via the chameleon mechanism.

The fifth force is given by
\begin{equation}
F_{\varphi} = -\frac{1}{2}\nabla f_R
\label{eq:fifth_force_fofr}
\end{equation}
where $f_R\equiv \dd f(R)/\dd R$ is the scalar field. In a FLRW background spacetime and in the quasi-static limit \citep{2014PhRvD..89b3521N}, the field-equation becomes 
\begin{align}\label{eq:fofr_full}
&\nabla^2 f_R = -\Omega_m H_0^2 a^2\left(\frac{\rho_m}{\overline{\rho_m}} - 1\right)+ a^2H_0^2\Omega_m\times \nonumber\\
&\left[\left(1 + 4\frac{\Omega_\Lambda}{\Omega_m}\right)\left(\frac{f_{R0}}{f_R}\right)^{n+1} - a^{-3} - 4\frac{\Omega_\Lambda}{\Omega_m}\right],
\end{align}
where $n$ and $f_{R0}$ are dimensionless model parameters.
The maximum enhancement of the gravitational force, i.e. the enhancement when there is no screening, is simply
\begin{equation}
\gamma_{\rm max} = 2 \beta^2 = \frac{1}{3}\;,\end{equation}
so gravity can be enhanced by up to $33\%$. How the screening mechanism works is explained in more detail in Sec.~\ref{sect:unified}.

\subsubsection{DGP}
\label{sec:ndgp}

The \citet*{2000PhLB..485..208D} (DGP) model is an example of a braneworld model where we are confined to live in a four-dimensional brane that itself is embedded in a five-dimensional spacetime. The DGP model can give rise to self-acceleration of the Universe without an explicit cosmological constant, but this branch has problems with instabilities. We  take the normal branch (n)DGP model as our main example. To get accelerated expansion in this branch, we need to add dark energy, and for simplicity, we assume that the background expansion is the same as in $\Lambda$CDM. This choice is not expected to change any of our conclusions.

Since it is a higher dimensional theory, the DGP model only fits into the formalism of Eq.~(\ref{eq:action}) as an effective theory. However, the modifications of gravity in this model can also be described as a fifth force given by
\begin{align}
F_{\varphi} = \frac{1}{2}\nabla\varphi,
\end{align}
where the evolution of the scalar field $\varphi$, the co-called brane-bending mode, is determined by the field equation\footnote{For the self-accelerating branch, the same expression holds but with $r_c\to -r_c$. In this case, $H(a)/H_0 = \frac{1}{2r_cH_0} + \sqrt{\Omega_m a^{-3} + (4r_c^2H_0^2)^{-1}}$ with $\Omega_m = 1-\frac{1}{r_cH_0}$.}
\begin{align}
\nabla^2\varphi + \frac{r_c^2}{3\beta_{\rm DGP}(a)a^2} \left((\nabla\varphi)^2 - (\nabla_i\nabla_j\varphi)^2\right) = \frac{\Omega_m H_0^2}{a\beta_{\rm DGP}(a)}\delta_m
\label{eq:nablasqphi_mabetaa}
\end{align}
where
\begin{align}
\beta_{\rm DGP}(a) = 1 + 2(r_cH_0) \frac{H(a)}{H_0}\left(1  + \frac{\dot{H}}{3H^2}\right).
\end{align}
The maximum value of the gravitational force in this model is 
\begin{align}
\gamma_{\rm max} = \frac{1}{3\beta_{\rm DGP}}.
\end{align}
For high values of $r_c$, we have $\beta_{\rm DGP} (a=1) \propto r_c$, so $\gamma_{\rm max} \propto \frac{1}{r_c}.$ The larger $r_c$,  the weaker the effect of the fifth force.

As also stated in \citet{2010PhRvD..81j3002S}, in spherical symmetry we can integrate Eq.~\eqref{eq:nablasqphi_mabetaa} directly to obtain
\begin{align}
\frac{\varphi'}{r} + \frac{2r_c^2}{3\beta_{\rm DGP}(a)a^2} \left(\frac{\varphi'}{r}\right)^2 = \frac{\Omega_m H_0^2}{a\beta_{\rm DGP}(a)}\frac{\int_0^r \delta_m r^2 \dd r}{r^3}.
\label{eq:ndgp_fieldeq_sphere}
\end{align}
This translates to a solution for the fifth force profile given by
\begin{align}
\gamma =\frac{\frac{1}{2}\varphi'}{\Phi'}=  \frac{1}{3\beta_{\rm DGP}}\left[\frac{2(-1 + \sqrt{1 + \epsilon(r)})}{\epsilon(r)}\right]
\label{eq:ndgp_fifth_force_sphere}
\end{align}
where
\begin{align}
\epsilon(r) = \frac{8(r_cH_0)^2\Omega_m}{3\beta_{\rm DGP}^2(a)a^3}\frac{\int_0^r \delta_m r^2 \dd r}{r^3}.
\end{align}
We have two regimes. For $\epsilon \ll 1$ (large $r$), we get $\gamma =  \frac{1}{3\beta_{\rm DGP}}$ and gravity is maximally modified. On the other hand, for $\epsilon \gg 1$ (small $r$), we have
\begin{align}
\gamma = \frac{1}{3\beta_{\rm DGP}} \frac{2}{\sqrt{\epsilon}} \ll  \frac{1}{3\beta_{\rm DGP}},
\end{align}
and the fifth-force is screened.

\subsection{The unified $\{m(a),\beta(a)\}$ description of chameleon-like models}\label{sect:unified}
\citet{2012PhRvD..86d4015B,2012PhLB..715...38B} show that any scalar-tensor theory that screens using the Chameleon and/or Symmetron mechanism can be described uniquely by the evolution of the mass and matter coupling along the cosmological attractor.
Thus parametrizing these two functions, which have a clear intuitive meaning -- the range and the strength of the fifth force, respectively --  to be effectively parametrized. The mapping between the potential $V(\varphi)$ and the coupling function $A(\varphi)$ to $\{m(a),\beta(a)\}$ has been derived in \cite{2012PhRvD..86d4015B} and reads as (in parametric form)
\begin{align}\label{eq:mapping}
V(a) =& V_0 - \frac{3}{M_{\rm Pl}^2}\int_{a_i}^a \frac{\beta^2(a)}{am^2(a)}\rho^2_m(a) \dd a\\
\varphi(a) =& \varphi_i + \frac{3}{M_{\rm Pl}}\int_{a_i}^a \frac{\beta(a)}{am^2(a)}\rho_m(a) \dd a\\
\log A(a) =& \log A(a_i) + \frac{3}{M_{\rm Pl}^2}\int_{a_i}^a\frac{\beta^2(a)}{am^2(a)}\rho_m(a)da,
\end{align}
where $\rho_m(a) = 3H_0^2M_{\rm Pl}^2\Omega_m/a^3$.

One of the advantages of using this form is the direct relationship between the evolution of linear matter perturbations and the $\beta$, $m$ functions
\begin{align}
\ddot{\delta}_m + 2H\dot{\delta}_m = \frac{3}{2}\Omega_m a^{-3} \delta_m \left(1 + \frac{2\beta^2(a) k^2}{k^2+ a^2m^2(a)}\right)\;.
\end{align} 
The two functions are, in the case of the Symmetron model, given by
\begin{align}
\frac{m(a)}{m_0} =& \frac{\beta(a)}{\beta_0} = \sqrt{1 - \left(\frac{a_{\rm ssb}}{a}\right)^3};
\end{align}
i.e., they follow the same evolution. The parameters $\beta_0$ and $m_0$ are related to the model parameters described in the previous section via $\beta_0 = \beta$ and $m_0 = \frac{1}{L}$.

For the Hu-Sawicki $f(R)$ model, on the other hand, we have
\begin{align}
\beta(a) &= \frac{1}{\sqrt{6}}
\end{align}
and
\begin{align}
m(a) &= \frac{H_0 \sqrt{\Omega_m + 4\Omega_\Lambda} }{\sqrt{|f_{R0}|}(n+1)}\left(\frac{\Omega_m a^{-3} + 4\Omega_\Lambda}{\Omega_m + 4\Omega_\Lambda}\right)^{n/2+2},
\end{align}
which means that $\beta$ is constant, and $m(a) \propto a^{-3(n/2+1)}$ for small $a$.
Note that the nDGP model does not map on to this parametrization except for the case of linear perturbations where it would correspond to $m(a) = 0$ and $\beta(a) = 1/\sqrt{6\beta_{\rm DGP}}$.

When the scalar field, in a region of density $\rho$ is close to the minimum of the effective potential (this corresponds to the fully screened regime), we have $\varphi \approx \varphi(a_\rho)$ where $a_\rho = a_{\rm env}(\overline{\rho_m}/\rho)^{1/3}$. The screening condition for an object with Newtonian potential $\Phi_N$ in a region of density $\rho$ can therefore be written as\footnote{This condition is derived under the assumption of a spherical top-hat overdensity.}
\begin{align}\label{eq:screencond}
\epsilon(a_\rho,a_{\rm env}) = \text{min}\left[\frac{|\varphi(a_\rho)-\varphi(a_{\rm env})|}{2\beta(a_{\rm env})M_{\rm Pl}\Phi_N},1\right].
\end{align}
An object is screened whenever $\epsilon(a_\rho,a_{\rm env}) \ll 1$. Here, $a_{\rm env} = a(\overline{\rho_m}/\rho_{\rm env})^{1/3}$ where $\rho_{\rm env}$ is the density of the environment the object is located in. The $\min$ condition ensures that we get the correct value, $\epsilon = 1$, in the nonscreened regime. The screening condition is related to $\gamma$ via
\begin{align}
\gamma = 2\beta^2(a_{\rm env})\epsilon(a_\rho,a_{\rm env}).
\end{align}
We should note that this is a very rough analytical approximation, but it is usually good enough for order-of-magnitude estimates. To get accurate predictions, we need to solve the field equation numerically, as discussed in the next section. 

\subsection{Numerical solving methods}
\label{sec:numerical_solving_methods}
To obtain accurate predictions for the field and force profiles in the models we study we need to numerically solve the field equations Eqs.~(\ref{eq:symm_full}) and~(\ref{eq:fofr_full}).
We solve these equations by discretizing them on a grid, using the NFW density profile for $\rho_m$ (see Appendix~\ref{sec:nfw-profile} for the relevant NFW-equations), and then use Newton-Gauss Seidel relaxation with multigrid acceleration to obtain the solution. 

In spherical symmetry the field equation becomes one-dimensional, $\phi = \phi(r)$ and $\nabla^2\phi = \frac{d^2\phi}{dr^2} + \frac{2}{r}\frac{d\phi}{dr}$. For the Symmetron model, we implemented a simple fixed-grid multigrid solver for this one-dimensional problem. To do this we write Eq.~(\ref{eq:symm_full}) as $\mathcal{L}_{i} = 0$ where
\begin{align}
\mathcal{L}_{i} = (\nabla^2\varphi)_i - \frac{a^2}{2\lambda_0^2}\left(\varphi_i\frac{a_{\rm ssb}^3}{a^3}\frac{\rho_m(r_i)}{\overline{\rho_m}} - \varphi_i + \varphi_i^3 \right)
\end{align}
where subscript $i$ denotes the value at the $i$'th gridpoint and
\begin{align}
(\nabla^2\varphi)_i = \frac{\varphi_{i+1}+\varphi_{i-1}-2\varphi_i}{(\Delta r)^2} + \frac{2}{r_i}\frac{\varphi_{i+1}-\varphi_{i-1}}{\Delta r}
\end{align}
is a second-order discretization of the Laplacian. We start by making a guess for the solution, and then we loop through the grid updating the solution using 
\begin{align}
\varphi_i \leftarrow \varphi_i - \frac{\mathcal{L}_i}{\partial \mathcal{L}_i/\partial \varphi_i}. 
\end{align}
The multigrid technique is used to speed up convergence. For more details about the methods we have used see \cite{Llinares2014a}.

The $f(R)$ equation Eq.~(\ref{eq:fofr_full}) is stiffer than the Symmetron equation for low values of $|f_{R0}|$, and a simple fixed-grid one-dimensional solver does not converge for the whole range of considered parameters.
The main reason for this is that $|f_{R}|$ is required to be strictly larger than zero (but can get very close).
This means that one has to ensure that $|f_{R}|$ does not become negative in the solving process. To achieve this, we used the already well-tested $f(R)$-module in the \texttt{ISIS} code \citep{Llinares2014a}. The only modification to the module is the change to a purely geometrical refinement criterion based on the value of $\rho_m$.
In total, we used $10-15$ levels of refinements starting from a base grid of $N=64^3$ grid nodes in order to get accurate field profiles down to very small radii. The Symmetron is also implemented in the {\tt{ISIS}} code, and we tested that our Symmetron field profiles agree very well with those obtained with our code.

After having calculated the field profiles, the fifth force can be calculated by using Eqs.~\eqref{eq:fifth_force_symm} and~\eqref{eq:fifth_force_fofr} for the Symmetron and $f(R)$ model, respectively.
For nDGP the analytical solutions to the field equation and force profile in spherical symmetry is given by Eqs.~\eqref{eq:ndgp_fieldeq_sphere} and~\eqref{eq:ndgp_fifth_force_sphere}, respectively. This makes it needless to solve any differential equations numerically\footnote{Generally, this is true for all models that have a derivative shift symmetry (like Galileons) and where the field equation is second order. If this is satisfied, then the equation of motion can be integrated to yield an algebraic equation in $\frac{\dd\varphi}{\dd r }\propto F_\varphi$}. The derivation of the analytical solution is shown in Appendix~\ref{sec:dgp_nfw}.

\begin{figure*}
  \centering
  \includegraphics[width=\textwidth]{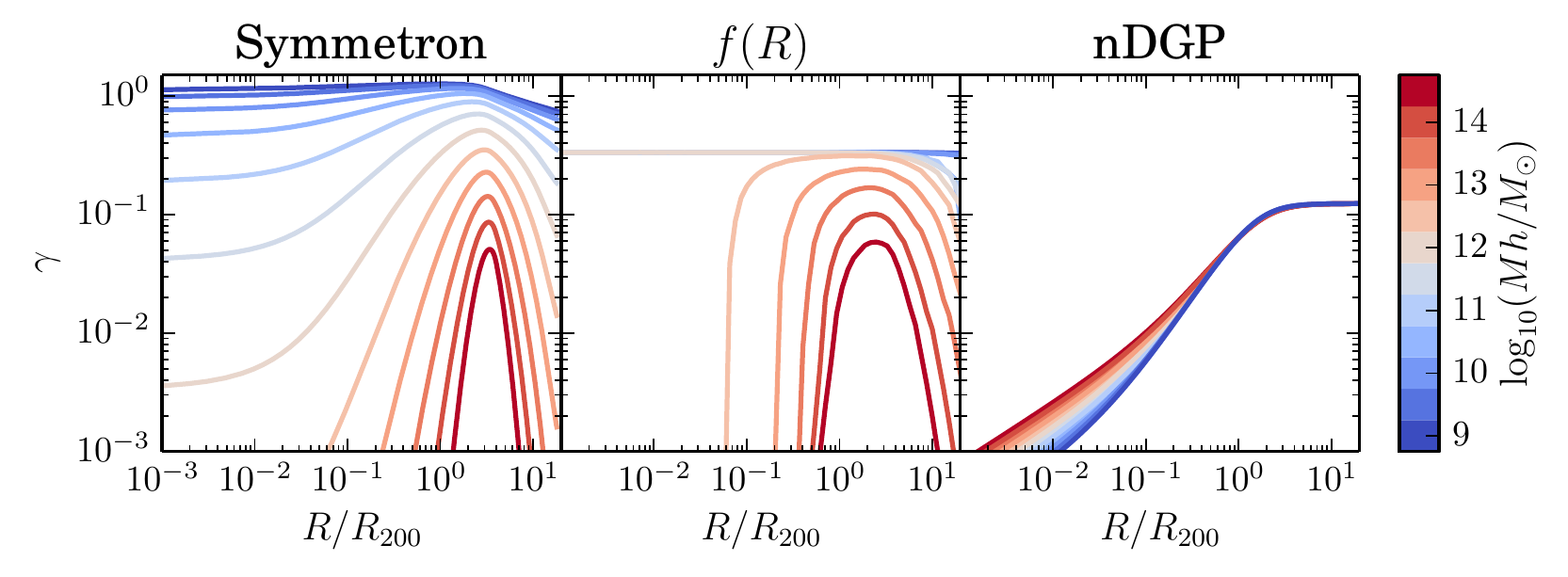}
  \caption{$\gamma$ vs $R/R_{200}$ for NFW halos with masses ranging from $10^{9}\Msh$ to $10^{14.5}\Msh$. The \textit{left panel} shows the Symmetron solution using $(a_{\rm ssb},\, L,\, \beta) = (0.7,\, 1\Mpch,\, 1)$, the \textit{central panel} the \citeauthor*{Hu2007} $f(R)$ solution with $(|f_{R0}|,\, n) = (10^{-6},\,1),$ and the \textit{right panel} the nDGP solution with $r_c = 1$.}
  \label{fig:gamma_vs_R}
\end{figure*}

\begin{figure*}
  \centering

\includegraphics[width=0.8\columnwidth]{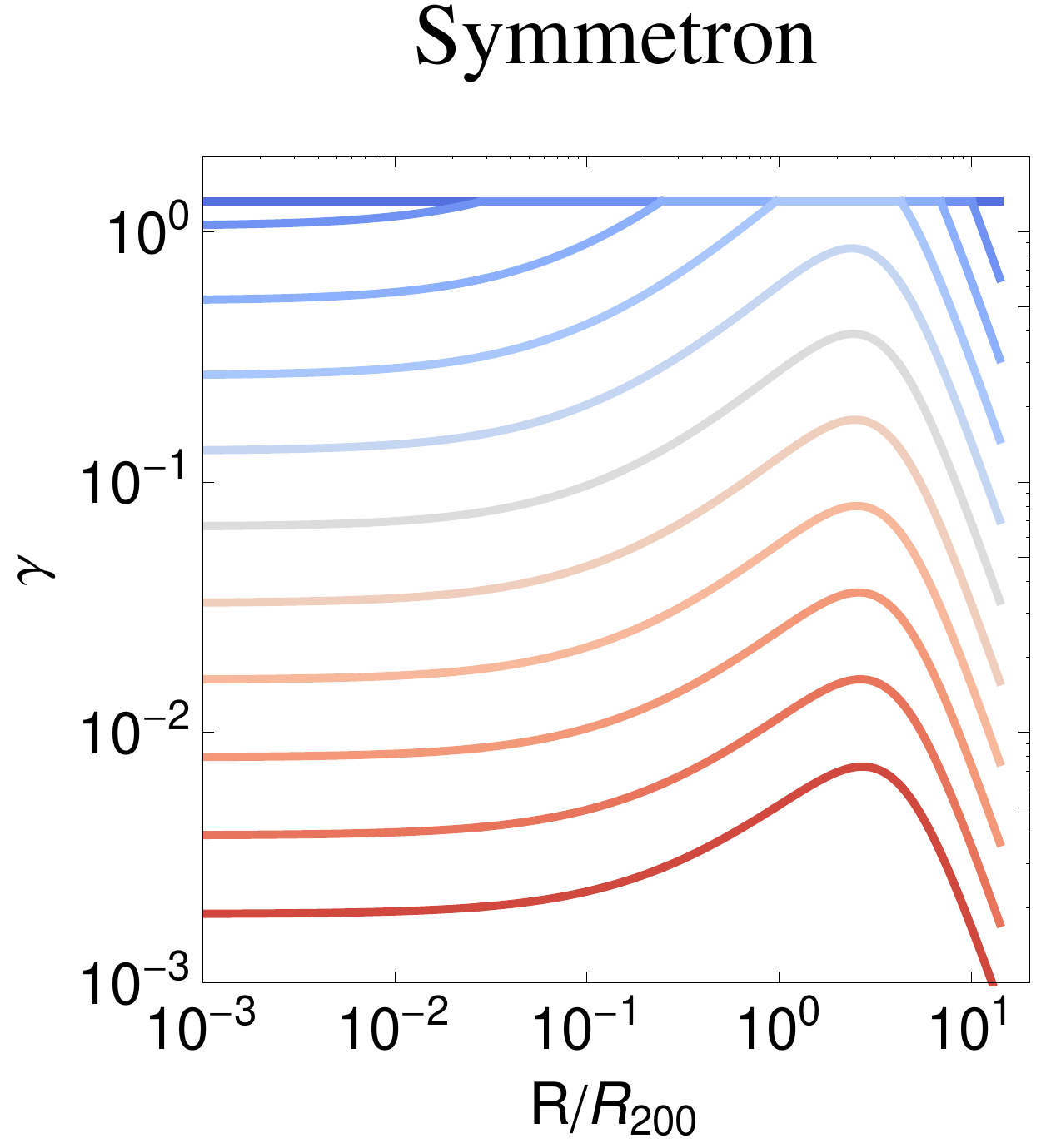}
    \includegraphics[width=0.8\columnwidth]{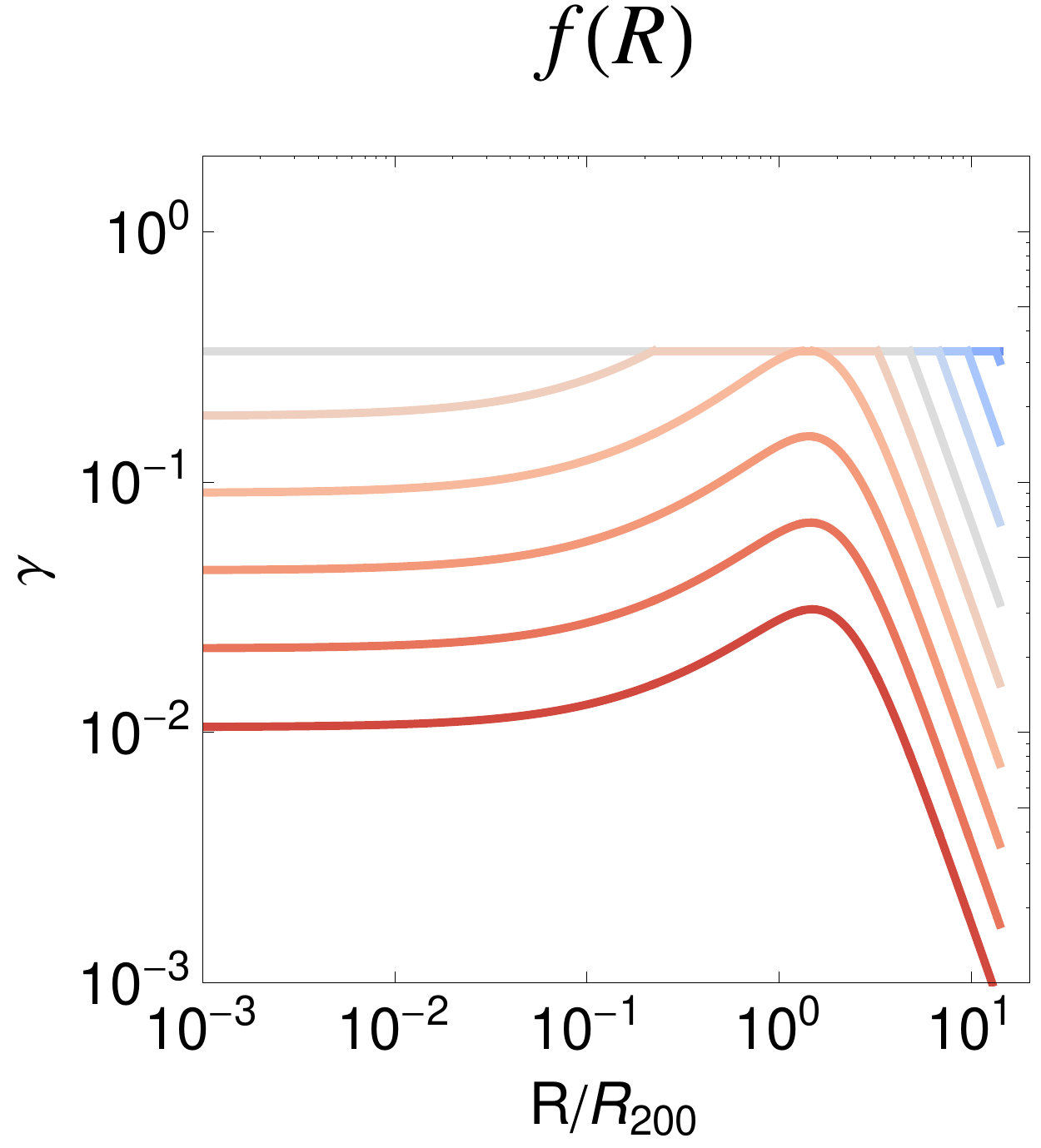}
  \caption{Semi-analytical solutions for $\gamma$ for the Symmetron (left) and $f(R)$ (right) using the same parameter values (and color coding) as in Fig.~\ref{fig:gamma_vs_R}.}
  \label{fig:gamma_analytic}
\end{figure*}

\section{Results}
\label{sec:results}
In this section we present our results from numerical solutions of the field equations (\S\ref{sec:force-profiles}, \S\ref{sec:gammabar}) and the obtained rescaling method for the three example models. The rescaling is first found empirically in \S\ref{sec:rescaling} and then semi-analytically in the $\{m(a),\,\beta(a)\}$ formulations in \S\ref{sec:rescaling_ma-betaa}.

\begin{figure*}
  \centering
  \includegraphics[width=\textwidth]{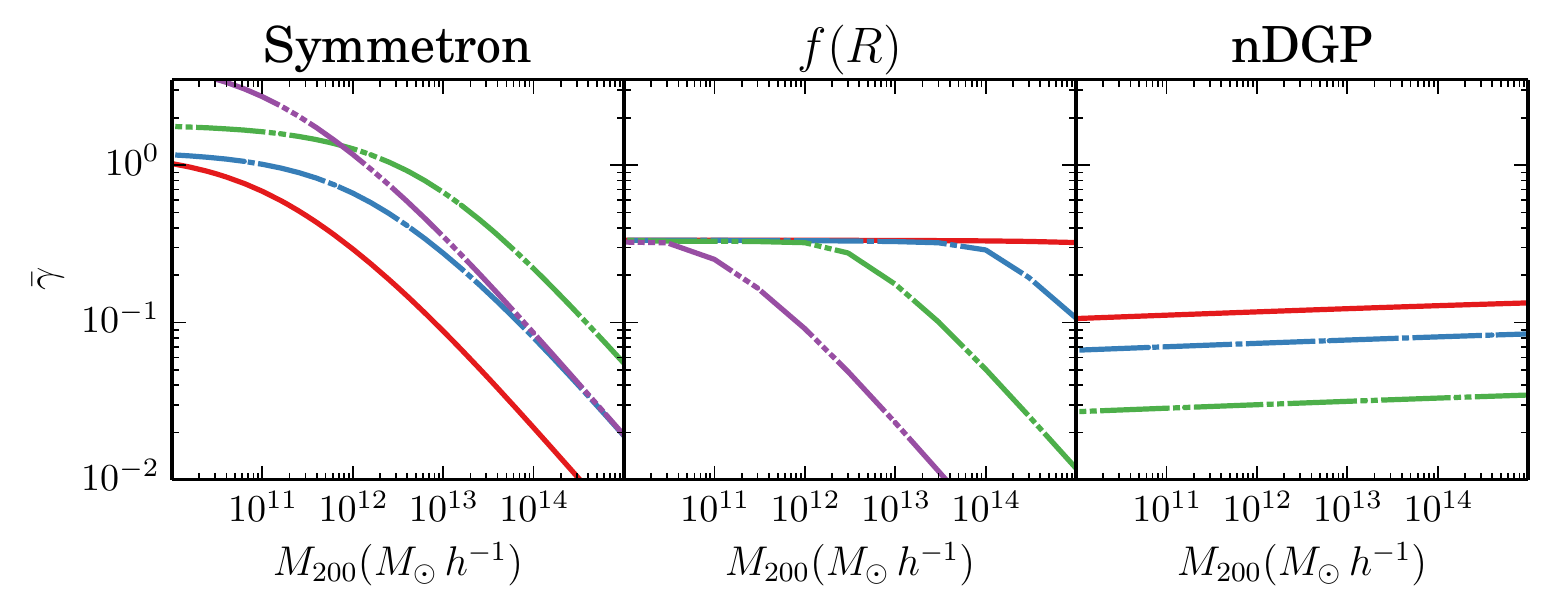}
  \caption{$\bar\gamma$ vs halo mass $M$. The \textit{left panel} shows the Symmetron curves with $(a_{\rm ssb},\, L,\, \beta)$ being (from solid to more interceptions) $0.7,\, 1\Mpch,\, 1$ (red), $0.7,\, 2\Mpch,\, 1$ (blue), $0.3,\, 1\Mpch,\, 1$ (green), $0.7,\, 1\Mpch,\, 2$ (purple). The \textit{central panel} shows the \citeauthor*{Hu2007} $f(R)$ solution with $\log_{10}|f_{R0}| = (-4,\,-5,\,-6,\,-7)$ (in red, blue, green, and purple, respectively) and the \textit{right panel} the nDGP solution with $r_c = (0.5,\,1\,,3)$ (from top to bottom).}
  \label{fig:gammabar}
\end{figure*}

\subsection{Force profiles}
\label{sec:force-profiles}
Figure~\ref{fig:gamma_vs_R} shows the radial profiles of $\gamma$  obtained solving Eqs.~(\eqref{eq:symm_full},~\ref{eq:fofr_full},~\ref{eq:ndgp_fifth_force_sphere}) for the Symmetron, $f(R)$, and, nDGP model, respectively (from left to right panels). We display a set of halo masses for each model using fixed model parameters.

The differences in shape between the models are striking. The Symmetron model shows a whole range of solutions: from fully screened in the center for the heaviest clusters to not screened at all for the lightest ones. The maximum of the fifth force seems to be located $\sim 1\--10R_{200}$ and drops again for larger radii. This drop happens due to the fifth force having a finite range, $F_\phi \propto \frac{1}{r^2}e^{-mr}$.

The $f(R)$ model presented in the central panel of Fig.~\ref{fig:gamma_vs_R} inherits a much more abrupt change from fully screened to not screened in the central halo region. With this particular choice of parameters ($|f_{R0}| = 10^{-6},\,n=1$), halos with a mass of $\gtrsim 10^{12} M_\odot$ seem to be fully screened in the central ($R\lesssim 0.1 R_{200}$) region, whereas in lighter halos, the fifth force stays at its theoretical maximum $1/3 F_{N}$. For comparison, in Fig.~\ref{fig:gamma_analytic} we show the force profiles derived using the semi-analytical method described in \S\ref{sect:semianal} for the $f(R)$ (left) and Symmetron (right panel) models. The agreement is fairly good around the most significant region $R\sim 1-5 R_{200}$ where $\gamma$ peaks. The semi-analytic results are presented in more detail in \S\ref{sect:semianal}.

In the righthand panel of Fig.~\ref{fig:gamma_vs_R}, we show the radial scalar field profile $\varphi(r)$ for a NFW halo in the nDGP model with $r_c = 1$. As is also possible to see from Eq.~\eqref{eq:ngdp_nfw_gamma}, the solution includes a curious feature: owing to the concentration-mass relation used (Eq.~\eqref{eq:NFW_c-M_relation}), heavier halos are less screened in the central region of the halo. Generally, the fifth force is less strong in the nDGP model and follows a similar functional form that is almost  independent of mass\footnote{The only mass dependence is implicit in that NFW halos of different sizes have different values for the concentration $c(M)$.}.

\subsection{Profile variation with halo mass}
\label{sec:gammabar}
To capture the change of the fifth force profile with halo mass, we chose to show the \textit{\emph{mass-weighted mean of the fifth force}} 
\begin{equation}
\bar\gamma = \int_0^{x_{\rm cut}}\dd x \gamma(x) w(x)
\label{eq:gammabar_def}
\end{equation}
where $w(x)$ is the normalized mass fraction $w(x)\dd x \equiv \dd M / M(<x_{\rm cut})$ as defined in Eq.~\eqref{eq:dMnfw}. This quantity was introduced by \citet{2010PhRvD..81j3002S} for the $f(R)$ and DGP model and also considered in \cite{2012JCAP...01..030C} for the symmetron model. The outer cutoff was fixed to $x_{\rm cut} = 10$. This value was chosen so that $w(x_{\rm cut})$ is roughly one-tenth of the maximum weight, meaning that not too much information is lost. As discussed in \S~\ref{sec:caveats} (see Eq.~\eqref{eq:gamma_masses}), $\bar\gamma$ can be\textbf{a} related to the ratio of the dynamical mass to the lensing mass.

Figure~\ref{fig:gammabar} shows $\bar\gamma$ versus the halo mass. However, this measure is not unique. Instead it is also possible, for example, to use a difference between the force in the center and in the outskirts of the halo. We discuss the advantages and caveats of using $\bar\gamma$ as an ``observable'' in \S\ref{sec:caveats}.

The lefthand panel of Fig.~\ref{fig:gammabar} shows three Symmetron solutions with different model parameters $(a_{\rm ssb},\,L,\,\beta)$. The slope, as well as the offset, of the three curves is seen to be fairly different. Also notable is the immense change in $\bar\gamma$ with changing halo mass. The $f(R)$ solutions (shown in the central panel of Fig.~\ref{fig:gammabar}) possess an upper limit at the already mentioned value of one-third. For lighter halos $\bar\gamma$ falls off rapidly. In the righter-most panel of Fig.~\ref{fig:gammabar}, we show the variation in $\bar\gamma$ with mass for the nDGP model for three different values of $r_c$. Overall, the impact of the fifth force is much weaker than in the other models. As discussed in \S\ref{sec:force-profiles}, the variation in the fifth force with halo mass is only due to the mass-concentration relation of the NFW halo. This feature manifests itself in Fig.~\ref{fig:gammabar} in the slight increase in $\bar\gamma$ with increasing halo mass.

\subsection{Empirical rescaling}
\label{sec:rescaling}
To find a rescaling that maps the various characteristics presented in the last section onto one single curve, we need to rescale the $x$ and the $y$ axes of Fig.~\ref{fig:gammabar}. The latter is naturally normalized to the theoretical possible maximum of the gravitational enhancement $\gamma_{\rm max}$. The rescaling of the halo mass is less trivial. Therefore, we simply define a characteristic mass $\mu_{200}$ in the following way:
\begin{equation}
\bar\gamma(\mu_{200}) \equiv \frac{1}{2}\gamma_{\rm max}
\label{eq:mu200_def}
\end{equation}
where $\bar\gamma$ is still the mass-weighted average of the enhancement of the fifth force as defined in Eq.~\eqref{eq:gammabar_def}
In general $\mu_{200}$ is a function of the various model parameters. Halos with mass much higher than $\mu_{200}$ are in the fully screened regime, whereas halos with mass much lower than $\mu_{200}$ are in the unscreened regime.

In this section we find the functional form of $\mu_{200}$ for each model merely empirically from our numerical solutions. However, in the next section, we derive semi-analytic predictions for a large class of scalar tensor theories.

This means that, by looking at Fig.~\ref{fig:gammabar}, it is possible to read off several values of $\mu_{200}$. For example, for the $f(R)$ model with $(n,\,|f_{R0}|)=(1,\,10^{-6})$ (in green), $\gamma(\sim 10^{13}M_\odot h^{-1}) = 1/6 = 1/2 \gamma_{\rm max}$, i.e., $\mu_{200}$, is in that particular case $\sim 10^{13}M_\odot h^{-1}$.

\begin{figure}
  \centering
  \includegraphics[width=.8\linewidth]{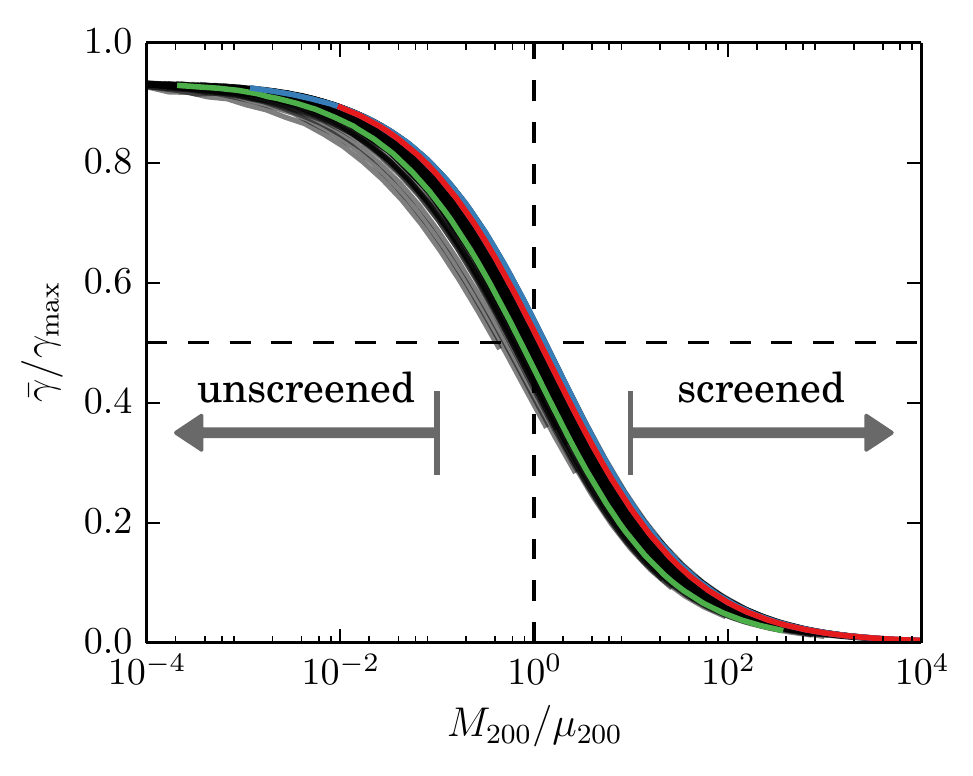}
  \caption{$\bar\gamma$ versus $M / \mu_{200}$ for the Symmetron model. The solid colored lines indicate the same model parameters as in Fig.~\ref{fig:gammabar}. The gray lines are an additional $124$ randomly chosen Symmetron parameters. $\mu_{200}$ for the Symmetron is given by Eq.~\eqref{eq:mu200_symm}. The black dashed lines indicate the point defining $\mu_{200}$ in Eq.~\eqref{eq:mu200_def}, and the gray arrows point to the screened and unscreened regimes.
}
  \label{fig:gammabar_vs_mu_rescaled_symm}
\end{figure}

For the Symmetron model, we found $\mu_{200}$ to be well fit by
\begin{equation}
\mu_{200}^{\rm Symmetron} = 2\times 10^{10} M_\odot h^{-1} \times \left(\frac{L}{\Mpch} \right)^3 a_{\rm ssb}^{-4.5}
\label{eq:mu200_symm}
,\end{equation}
and $\gamma_{\rm max}$ is given by Eq.~\eqref{eq:gammamax_symm}.
Figure~\ref{fig:gammabar_vs_mu_rescaled_symm} shows the resulting rescaled curves, i.e., $\bar\gamma / \gamma_{\rm max}$ versus $M/\mu_{200}$. In addition to the four parameter sets already presented in the lefthand panel of Fig.~\ref{fig:gammabar} (keeping the color coding), we plot an additional $125$ curves with randomly chosen parameters. Specifically, we drew $L\sim [0.2,\,2]$\Mpch, $a_{\rm ssb}\sim [0.2,\,0.8]$, and $\beta\sim [0.2,\,2.4]$ (all uniformly).

It is notable that in spite of this wide parameter range, the resulting curves in Fig.~\ref{fig:gammabar_vs_mu_rescaled_symm} are very similar. All follow the characteristic shape -- from $\gamma \sim \gamma_{\rm max}$ at $M \ll \mu_{200}$ through $\gamma(M\approx\mu_{200}) = \gamma_{\rm max}/2$ to $\gamma \sim 0$ for $M\gg\mu_{200}$ -- with very little scatter. We note, however, that for low masses, $\bar\gamma$ does not fully approach $\gamma_{\rm max}$ but falls $\sim 10\%$ short. This is due to the drop in $\gamma$ for larger radii as is visible in the lefthand panel of Fig.~\ref{fig:gamma_vs_R}.

\begin{figure}
  \centering
  \includegraphics[width=.8\linewidth]{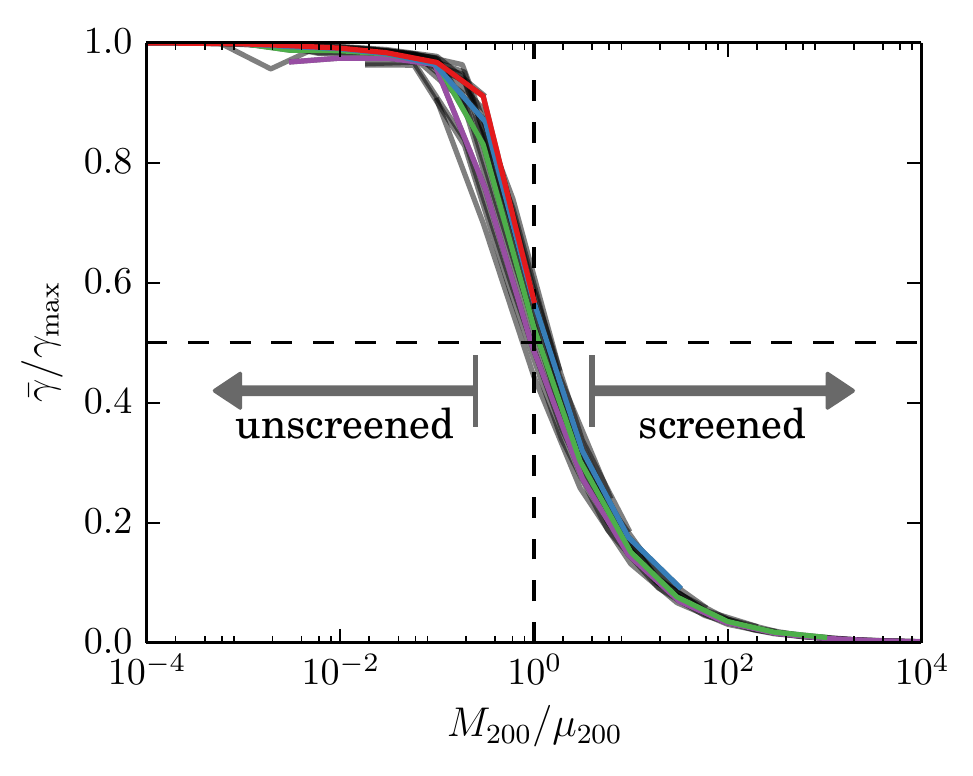}
  \caption{$\bar\gamma$ versus $M / \mu_{200}$ for the Hu-Sawicki $f(R)$ model. The solid colored lines indicate the same model parameters as in Fig.~\ref{fig:gammabar}. The gray lines are an additional $5$ values of $f_{R0}$. $\mu_{200}$ for this model is given by Eq.~\eqref{eq:mu200_fofr}.}
  \label{fig:gammabar_vs_mu_rescaled_fofr}
\end{figure}

In Fig.~\ref{fig:gammabar_vs_mu_rescaled_fofr} we present the results of the same rescaling technique applied to the $f(R)$ results. In this case the rescaling mass is found to be
\begin{equation}
\mu_{200}^{f(R)} = 10^{13}M_\odot h^{-1}\times \left(\frac{|f_{R0}|}{10^{-6}}\right)^{1.5}
\label{eq:mu200_fofr}
\end{equation}
and $\gamma_{\rm max} = 1/3$.
The curves in Fig.~\ref{fig:gammabar_vs_mu_rescaled_fofr} correspond to the numerical solutions for each combination of $n=(1,\,2)$ and $|f_{R0}|=(10^{-8},\,3\times 10^{-8},\,10^{-7},\,3\times 10^{-7},\ldots,10^{-4}),$ with the ones already presented in Fig.~\ref{fig:gammabar} color-coded accordingly. Also here, the scatter around the mean form is quite small. The second striking feature the similarity in the overall shape of the curves in Figs.~\ref{fig:gammabar_vs_mu_rescaled_symm} and~\ref{fig:gammabar_vs_mu_rescaled_fofr} -- even though the solution stems from two independent modified gravity theories. The main differences between the two models are the slightly different transitions widths and, the fact that for $M\ll \mu_{200}$, $\gamma\approx \gamma_{\rm max}$ in the $f(R)$ case, whereas the Symmetron curves are limited to a lower value.

\begin{figure}
  \centering
  \includegraphics[width=.8\linewidth]{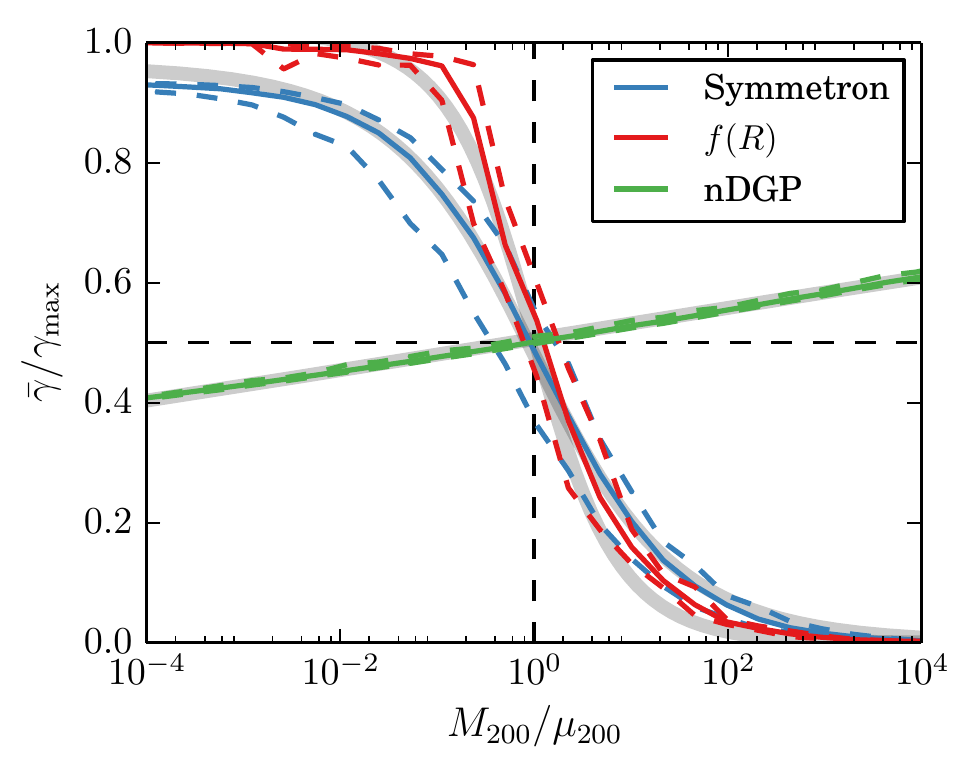}
  \caption{$\bar\gamma$ versus $M / \mu_{200}$ for the three example models. The solid [dashed] lines show the mean [minimum \& maximum] values for the $128$ (Symmetron), $16$ ($f(R)$), and $20$ (nDGP) numerical curves obtained. The solid gray lines show the fits of Eq.~\eqref{eq:gammabar_fit_general} described in \S\ref{sec:rescaling}.}
  \label{fig:gammabar_vs_mu_rescaled_means}
\end{figure}

Rescaling the $\bar\gamma$-mass relation for the nDGP model (presented in the right panel of Fig.~\ref{fig:gammabar}) is not as straightforward. Since the curves are nearly constant (and do not possess a characteristic drop), no empirical formulation for $\mu_{200}$ is apparent.
However, since calculating $\bar\gamma$ in the nDGP model is not computationally expensive, we can numerically solve Eq.~\eqref{eq:mu200_def} for several values of $r_c$. This let us fit the empirically found relationship
\begin{equation}
\log_{10}\mu_{200} = A - (r_c / a)^{b} {\rm e}^{-r_c / a}
,\end{equation}
which yields $(A,\,a,\,b)=(11.14,\,6.72,\,-0.57)$. The rescaled $\bar\gamma(M_{200})$ solutions obtained with $r_c = (0.5,\,1,\,\ldots,\,20)$ showed so little variation that we chose not to explicitly show them. Instead, we display in Fig.~\ref{fig:gammabar_vs_mu_rescaled_means} the average $\bar\gamma / \gamma_{\rm max}\--M_{200}/\mu_{200}$ curves (solid lines), as well as the maximal deviation found for each (dashed lines). Clearly, the nDGP curves show very little variation.

Figure~\ref{fig:gammabar_vs_mu_rescaled_means} once more shows the difference between the Symmetron and $f(R)$ model with their characteristic `inverse-S-shape' on the one side, and, the nDGP -- with a log-linear relation -- on the other.
A simple fitting function that is able to match the characteristic `inverse-S-shape' given by
\begin{align}\label{eq:fittingformula}
\bar{\gamma}(M) = \gamma_{\rm max}\left[1 - \frac{2}{\pi}\arctan(M_{200}/\mu_{200})\right.].
\end{align}
 However, it is also clear that the models of the former group do not show exactly identical behavior. Instead, the transition region for the Symmetron model is a bit wider, and, as mentioned before, it does not converge to $\gamma_{\rm max}$ for lower masses. This is why we propose the more general form
\begin{align}
\bar{\gamma}(M) = \mathcal{A}\gamma_{\rm max}\left(1 - \frac{2}{\pi}\arctan\left[(M_{200}/\mu_{200})^{\alpha}\right]\right)
\label{eq:gammabar_fit_general}
\end{align}
with two free parameters $\mathcal{A}$, $\alpha$. Here, the prefactor $\mathcal{A}$ can be interpreted as the part of the theoretical enhancement of gravity that is effectively active (and, thus, can only be $<1$), and $\alpha$ controls the width of the partially screened region with $\alpha = 0$ being `infinitely wide' and $\alpha\rightarrow\infty$ the transition turning into a step function. Also, if $\alpha<0$, the ``inverse-S-shape'' flips vertically and turns into an ``S-shape''. Quantitatively, we can relate $\alpha$ to the half width of the partially screened region through
\begin{equation}
W(\alpha) = \tan\left(\frac{\pi d}{2}\right)^{-1/|\alpha|}.
\label{eq:Walpha}
\end{equation}
Here, $M_{200}\in[\mu_{200}/W,\,W\mu_{200}]$ is the interval where  $d < \bar\gamma/\gamma_{\rm max} < \mathcal{A}(1 - d)$.

Figure~\ref{fig:gammabar_vs_mu_rescaled_means} also shows the results of fitting Eq.~\eqref{eq:gammabar_fit_general} to all of our numerical data (with enforcing $\mathcal{A}<1$). It is noteworthy that this simple, two-parameter fit describes the numerical results of the three studied models very well, even for the nDGP model. As expected, the $\alpha$ parameter reflects the fundamental difference between the nDGP and the two other models, as it is close to zero ($-0.03$) for the nDGP and more similar for the other two models. Specifically, we obtained $\alpha=0.46$ and $0.78$ for the Symmetron and $f(R)$ model, respectively. 

\begin{figure}
  \centering
  \includegraphics[width=.8\linewidth]{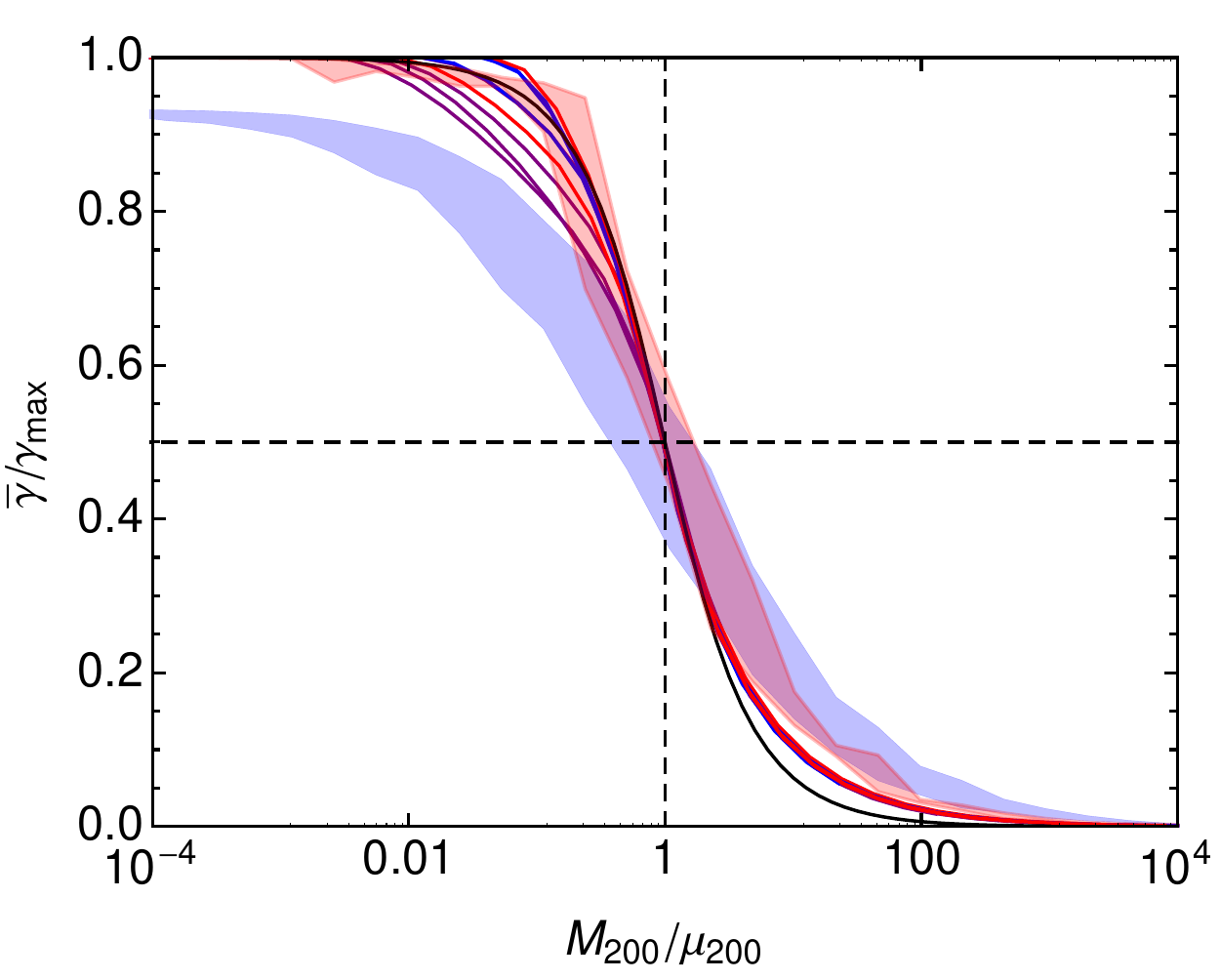}
  \caption{Semi-analytical predictions for $\bar\gamma$ versus $M / \mu_{200}$ for the Hu-Sawicki $f(R)$ model (red), the Symmetron (blue) and the Dilaton model (purple). The solid black curve shows the fitting formula Eq.~(\ref{eq:fittingformula}). The shaded regions show the results of the full numerical analysis.}
  \label{fig:analytic_gammabar_vs_mu_rescaled_fofr}
\end{figure}

\subsection{Semi-analytical derivation of $\mu_{200}$ in the $\{m(a),\beta(a)\}$ formulation}
\label{sect:semianal}
\label{sec:rescaling_ma-betaa}

In the following, we discuss the origin of the empirical rescalings found. Using the general framework of the $\{\beta(a),m(a)\}$ formulation described in \S\ref{sect:unified}, we can derive an analytic approximation for $\mu_{200}$. It turns out that we can do this quite generally, so our results below can be applied to {\it any} scalar-tensor theory with a Lagrangian $L = \frac{1}{2}(\partial\varphi)^2 + V(\varphi)$ and a conformal coupling to matter. We only focus on a very rough derivation, which can be seen as the foundation of a more precise treatment in the future.

Before we do this, we should check to what extent our semi-analytic approximation, which formally holds only for a spherical top-hat overdensity, is valid for NFW halos. In Fig.~\ref{fig:gamma_analytic} we show the semi-analytical force profiles that can be compared to Fig.~\ref{fig:gamma_vs_R}, which shows the true numerical results. The qualitative agreement is fairly good. We are able to match the shape around the peak, which is the significant part of the profile that will give rise to most of the signal in $\bar{\gamma}$, and the amplitude is off by no more than a factor of a few, which is good enough for our purposes.

As a first -- rather crude -- approximation, we assume that a halo is fully screened within a certain radius $x_{\rm screen}$. In the unscreened regime $x_{\rm screen}\rightarrow 0,$ but generally $0\lesssim x_{\rm screen}\lesssim 10$. Thus, we approximate $\gamma(x) \simeq \gamma_{\rm max}\theta(x_{\rm screen}-x)$ where $\theta$ is the Heaviside function.
This leads to a solution for the mass-weighted average of $\gamma$, which reads as\begin{align}
\bar{\gamma} \simeq \gamma_{\rm max} \frac{M(<x_{\rm screen})}{M(<x_{\rm cut})}\;.
\end{align}
Furthermore, using the definition of $\mu_{200}$, namely $\bar{\gamma}(\mu_{200}) = \frac{1}{2}\gamma_{\rm max}$, we require
\begin{align}
\frac{M(<x_{\rm screen})}{M(<x_{\rm cut})}\bigg|_{M_{200} = \mu_{200}} \simeq \frac{1}{2},
\end{align}
which requires $x_{\rm screen} \sim \mathcal{O}(1)$ for typical halo masses.

The dependence of $x_{\rm screen}$, hence of $\mu_{200}$, on the various SMG model parameters can be calculated using the screening condition Eq.~\eqref{eq:screencond}. In this case, the condition reads as
\begin{align}
\label{eq:mucondition}
\frac{|\varphi(a_\rho(x))-\varphi(a_{\rm env})|}{2\beta(a_{\rm env})M_{\rm Pl}\Phi_N(x_{\rm screen})} = 1\\
a_\rho(x) = a_{\rm env}\left(\frac{\overline{\rho_m}}{\rho(x_{\rm screen})}\right)^{1/3}.
\end{align}
With $\Phi_N(r) = \frac{GM(<r)}{r}$ and $M_{200} = 4\pi/3\Delta_{\rm vir}\Omega_m\rho_{c0} r^3 \to r\propto M^{1/3}$, we can easily calculate how $\mu_{200}$ scales with the model parameters in any given model. 

For the Symmetron we find the scaling $M(<x_{\rm screen})^{2/3} \propto \frac{L^2}{a_{\rm ssb}^3}$. As a result, $\mu_{200}$ scales as
\begin{align}
\mu_{200}^{\rm Symmetron} = (\sim 0.4 \pm 0.3)\times 10^{10}M_{\odot} h^{-1}\times\left(\frac{L}{a_{\rm ssb}^{3/2}\text{Mpc}/h}\right)^{3}
\end{align}
where the proportionality factor was calculated numerically.
This scaling is exactly the same as we found empirically in the previous section, but the proportionality factor is off by a factor of $\sim 3\--10$.

For the Hu-Sawicki $f(R)$ model, we find the scaling $\mu_{200} \propto |f_{R0}|^{3/2}C^{3/2(n+1)} M_{\odot}/h$, where $C$ is a constant of order one, dependent only on $x_{\rm screen}$ and $\Omega_m$. Numerically we find that the results are only weakly dependent on $n$: changing $n$ from $1$ to $5$ only gives us a factor $1.5$ offset in the value we find for $\mu_{200}$. Again the proportionality factor is found numerically and gives us
\begin{align}
\mu_{200}^{f(R)} = (\sim 1.1\pm 0.4)\times 10^{13}M_{\odot} h^{-1} \times \left(\frac{|f_{R0}|}{10^{-6}}\right)^{3/2},
\end{align}
which is almost spot on the empirical relation found in the previous section.
 
For a general $\{m(a),\beta(a)\}$ model, we see from Eq.~\eqref{eq:mapping} that we would (again very roughly) expect to have $\phi(a) \sim \frac{\beta(a)}{a^3m^2(a)}$. This in the condition Eq.~\eqref{eq:mucondition} means that we generally expect $\mu_{200} \propto \lambda_\phi^3(a)$ where $\lambda_\phi(a) = m^{-1}(a)$ is the range of the fifth force.

In Fig.~\ref{fig:analytic_gammabar_vs_mu_rescaled_fofr} we show the semi-analytical profiles for $\bar{\gamma}$ in terms of $M_{200}/\mu_{200}$ for $f(R)$ and the symmetron, together with the full numerical results. To demonstrate that this rescaling works in general, and not just for the two models presented here, we also calculated the $\bar{\gamma}$ predictions for the Dilaton model presented in \cite{2012JCAP...10..002B,2010PhRvD..82f3519B}. This model is characterized by
\begin{align}
m(a) &= m_0 a^{-r}\\
\beta(a) &= \beta_0 e^{\frac{s}{2r-3}(a^{2r-3}-1)} 
\end{align}
where $r,s,m_0$, and $\beta_0$ are free model parameters and where the evolution of $\beta$ are very different from our two example models. Figure~\ref{fig:analytic_gammabar_vs_mu_rescaled_fofr} shows that the rescaling also works perfectly for the Dilaton model.

We have not been able to get a prediction for the width of the partially screened region semi-analytically, so we leave this for future work.

\begin{figure}
  \centering
  \includegraphics[width=\linewidth]{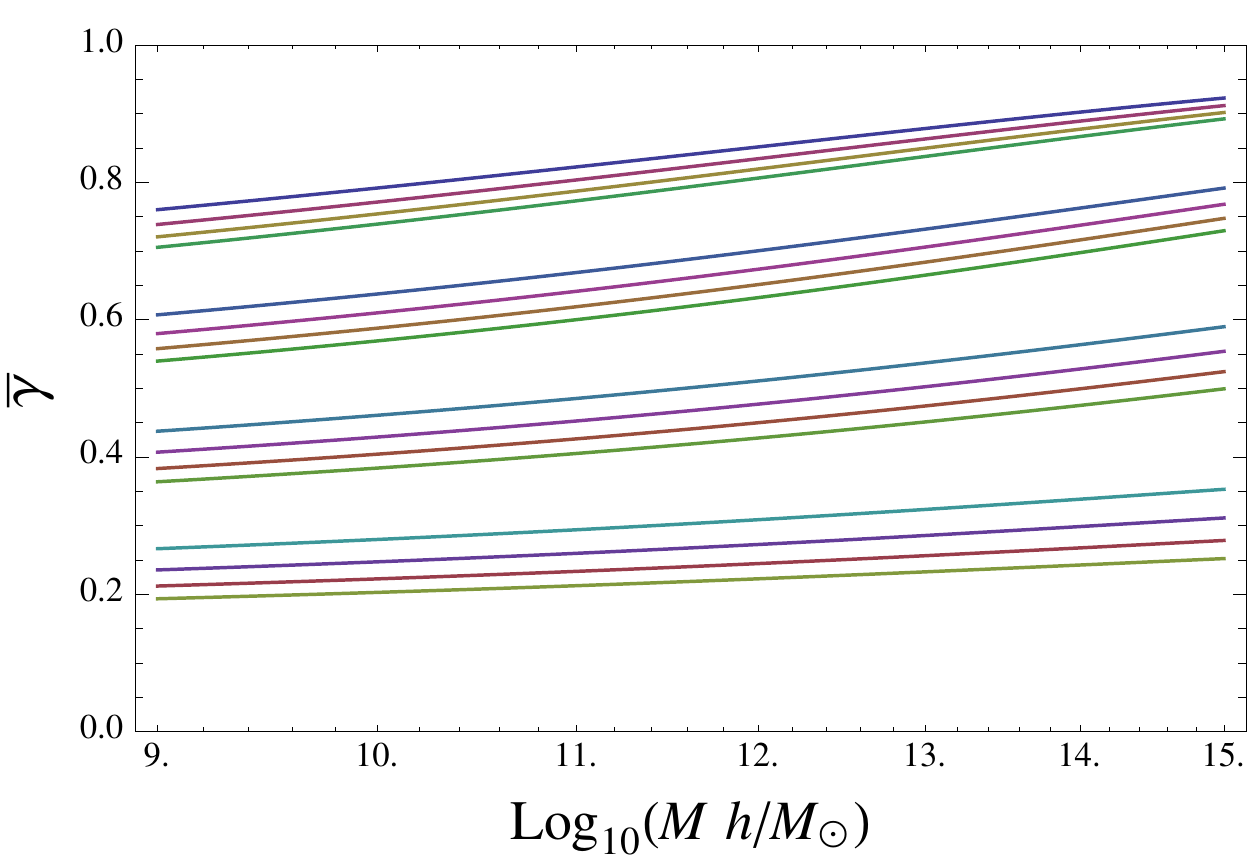}
  \caption{$\gamma(M)$ for NFW halos with masses ranging from $10^{9}\Msh$ to $10^{15}\Msh$ using Eq.~\eqref{eq:paramvs} with $G(z) = \frac{2((1+\zeta z)^n-1)}{\zeta z}$. The four bands of lines show the results (from top to bottom)  of $\zeta = 0.1,1.0,10,100,$ and within each band, we have (from top to bottom) $n = 0.5,0.4,0.3$,and $0.2$.}
  \label{fig:gamma_anal}
\end{figure}

\subsection{Predictions for other screening mechanisms}

For other screening mechanisms that are not encapsulated by the $\{m(a),\beta(a)\}$ formulation, it is harder to make general predictions. For any theory with a derivative shift symmetry (the Galileon symmetry), second-order equations of motion the field equation in spherical symmetry are integrable, leading to an algebraic equation on the form
\begin{align}
f(\phi'/r) = \frac{\int_0^r\delta_m r^2 dr}{r^3} = \frac{\Delta_{\rm vir}}{3x^3}\left(\frac{\log(1+cx)  - \frac{cx}{1+cx}}{\log(1+c) - \frac{c}{1+c}}\right) 
\end{align}
where the last fraction can be identified as $g_{\rm NFW}(c x) / g_{\rm NFW}(x)$ (see Eq.~\eqref{eq:gnfw}).
For a general model this leads to
\begin{align}\label{eq:paramvs}
\gamma = \gamma_{\rm max}G\left(\frac{g_{\rm NFW}(c)}{x^3g_{\rm NFW}(cx)}\right)
\end{align}
for some function $G$ that satisfies $G(0) = 1$ and $G(\infty) = 0$. For DGP we have $G(z) = \frac{2(\sqrt{1+\zeta z}-1)}{\zeta z}$ with $\zeta = \frac{8r_c^2H_0^2\Omega_m \Delta_{\rm vir}}{9\beta^2_{\rm DGP}}$.

For models that only have a shift symmetry, like the k-mouflage screening mechanism, the field equation,
\begin{align}
\frac{1}{r^2}\frac{d}{dr}[r^2f(X)\frac{d\phi}{dr}] = \frac{\beta\delta\rho_m}{M_{\rm Pl}},
\end{align}
can also be integrated up, and from this it follows that the fifth-force $F_\phi \propto \nabla\phi$ is a function of $\frac{\int_0^r\delta_m r^2 dr}{r^2}$ for spherical symmetry. This leads to a slightly different form
\begin{align}
\gamma = \gamma_{\rm max}F\left(\frac{g_{\rm NFW}(c)}{x^2g_{\rm NFW}(cx)}\right)
\end{align}
for some function $F$ satisfying $F(\infty) = 0$.

The exact form for the functions $F$ and $G$ are model dependent, making it hard to make definite statements here. However, one general feature we are able to deduce is that, just as we saw for DGP, $\gamma$ is only sensitive to the concentration of a halo. We therefore expect a fairly weak mass dependence of $\bar{\gamma}$ for models in this class.

As an example, we can try to parametrize the functional form for $F$ and $G$ and calculate $\bar{\gamma}$. One functional form that satisfies the requirements is $G(z) = \frac{(1+\zeta z)^{n}-1}{\zeta z}$ with $n\in (0,1)$. The results we get from this of course depend on the values we take for the parameters (like $\zeta$); however, taking realistic values for the parameters, which is motivated by the requirement of satisfying local gravity constraints, the typical result we find is as expected (see Fig.~\ref{fig:gamma_anal}): a very weak $M$ dependence of $\bar{\gamma}$.

\begin{figure}
\centering
\includegraphics[width=\linewidth]{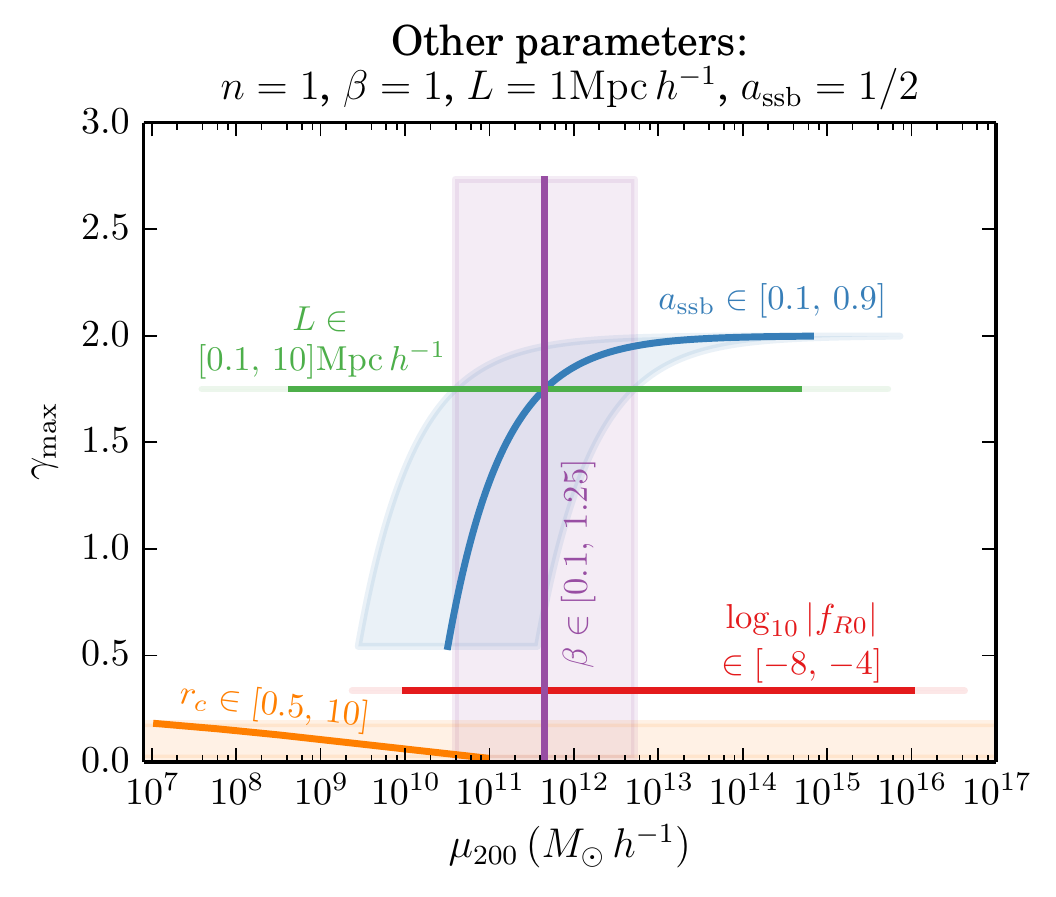}
\caption{Rescaling parameter $\mu_{200}$ versus the maximum gravitational enhancement $\gamma_{\rm max}$ for several discussed models (solid colored lines) and the respective partially screened regions (colored semi-transparent areas). The red line shows the space spanned by the Hu-Sawicki $f(R)$ model when varying $|f_{R0}|$ from $10^{-8}$ (left end) to $10^{-4}$ (right end), while leaving $n=1$ fixed. For the Symmetron, we also varied one parameter per curve (in blue, green, and violet), while leaving the others fixed at $\beta = 1$, $L = 1{\rm Mpc}\,h^{-1}$, $a_{\rm ssb}=1/2$. 
The orange line indicates the region for nDGP models with $r_c \in [0.5,\,10]$ (left to right) with the associated semi-transparent area spanning wider than the plot range.  See \S\ref{sec:observational_implications} for details.}
\label{fig:gamma-max_vs_mu-200}
\end{figure}

\section{Discussion}
\label{sec:discussion}

Our numerical force profiles presented in Sects. \ref{sec:force-profiles} and \ref{sec:gammabar} are in good agreement to what has been previously found in {\it N}-body simulations \citep[e.g.,][]{2011PhRvD..83d4007Z,2015arXiv150306673F}. The simple rescaling procedure based on the maximum enhancement of the gravitational force $\gamma_{\rm max}$ and the intermediately screened mass $\mu_{200}$ (\S\ref{sec:rescaling}) demonstrates that in spite of their foundational differences, the three screened-modified gravity models investigated show self-similar behavior in cluster environments.

\subsection{Observational implications}
\label{sec:observational_implications}
Our rescaling methods allow an abstract characterization of screened modified gravity theories, which themselves might possess quite different screening mechanisms. This potentially has interesting implications for constraining SMGs.

The colored lines in Fig.~\ref{fig:gamma-max_vs_mu-200} show $\gamma_{\rm max}$ versus $\mu_{200}$ for some selected models. 
The lines were drawn by altering one model parameter while leaving the others constant (see figure caption for details). Each set of SMG model parameters occupies a specific point on this graph; for example, the combination $(a_{\rm ssb},\,\beta,\,L) = (1/2,\,1,\,1\Mpch)$ maps to $(\gamma_{\rm max},\,\mu_{200})\approx(1.75,\,4.5\times 10^{11}M_\odot\,h^{-1})$. The semi-transparent areas in Fig.~\ref{fig:gamma-max_vs_mu-200} denote the extent of the partially screened region using Eq.~\eqref{eq:Walpha} with $d=0.2$. This means that for every selected value of $\mu_{200}$, the semi-transparent area indicates the range where the fifth force is at least $20\%$ but a maximum of $80\%$ active.

Thus, Fig.~\ref{fig:gamma-max_vs_mu-200} indicates \textit{(i)} what halos are partially screened and, \textit{(ii)} what the maximum enhancement of gravity is for this model. Physically, $\mu_{200}/W$ can be interpreted as the minimum mass where one expects a variation in halo properties; that is to say:
\begin{itemize}
\item For $M_{200}\lesssim \mu_{200}/W$ halos in this model are unscreened, thus allowing for comparison with the $\Lambda$CDM predictions. In particular, one expects that the majority of halo properties are rescaled by $\sim\gamma_{\rm max}$. This can make the comparison with $\Lambda$CDM difficult when using, for example, the dynamical mass as mass estimate.
\item Halos with $M_{200}\in [\mu_{200}/W,\,W\mu_{200}]$ are partially screened. This means that in this mass range, halo properties might differ from the expected. 
Since halos in this mass range are between screened and unscreened, the environment of these halos has a major effect \citep[see][for the study of this effect using {\it N}-body simulations]{2011PhRvL.107g1303Z,2012ApJ...756..166W}.
Also, comparing the inner and outer regions of individual objects can lead to insight (for instance, the ratio $M_{500}/M_{200}$). In Fig.~\ref{fig:gamma-max_vs_mu-200} the partially screened region is shown as a shaded area in the matching color. As demonstrated for the nDGP case, it is possible that -- for some SMGs -- \textit{\emph{all}} halos are always partially screened.
\item For masses $\gtrsim W\mu_{200}$, the halo is fully screened, so no deviation from the $\Lambda$CDM predictions is expected. 
\end{itemize}
Conclusively, in any observable-mass scaling relation, we predict a peculiar feature at $\mu_{200}$, where the observables match the $\Lambda$CDM prediction for higher masses and where they are different for lower masses. However, the latter statement only holds if the true mass of the halo is used. This is because mass tracers like the dynamical mass can also be affected by the fifth force \citep{2010PhRvD..81j3002S}.

The extent of the partially screened region, which we quantified with $\alpha$ (or, $W$, see Eq.~\eqref{eq:Walpha}, can potentially be used to distinguish between various screening mechanisms. We found that this extent is constant for each screened-modified gravity considered, but differs significantly between them. In particular, the value of $\alpha$ found in the numerical fits of \S\ref{sec:rescaling} translates to $\log_{10}W\sim 16$, $\sim 1,$ and $\sim 0.6$ for the nDGP, Symmetron, and the $f(R)$ models, respectively (using $d=0.2$, as above). In a $\gamma_{\rm max}$-$|\alpha|$ (or equivalently, a $\gamma_{\rm max}$-$W$) diagram, we expect eventual constraints to rule out the area $\gamma_{\rm max}>\gamma_{\rm constraint}$, $|\alpha|\lesssim \alpha_{\rm constraint}$ ($W\gtrsim W(\alpha_{\rm constraint})$).

\subsection{Caveats}\label{sec:caveats}
Several assumptions went into deriving the presented formalism. Here, we want to mention and investigate them.

\begin{itemize}
\item \textit{Halo profiles and shape}. Throughout this work, we assumed spherical symmetry and perfect NFW halos. However, in reality clusters of galaxies do not exactly follow an NFW mass distribution, and halos also possess a nonzero ellipticity \citep[e.g.,][]{Oguri2010MNRAS.405.2215O}. Another phenomenon falling in the same category is our assignment of a unique mass-concentration mapping, whereas the spread in this relation is fairly wide \citep{2007MNRAS.381.1450N}. Both are consequences of the fact that clusters are individual objects with their own histories and environments. As a result, individual clusters (as any astrophysical object) have mostly a limited explanatory power and must be stacked for analysis.
\item \textit{Model assumptions.} To rescale the $\bar\gamma\--\mu_{200}$ curves accordingly, an intermediate-mass scale has to exist where $\bar\gamma\sim\gamma_{\rm max}/2$. Since this is the scale between the screened and unscreened regimes and, our work addresses `screened modified gravity theories', we think that this can be safely assumed.
The universal rescaling works perfectly for the models we have tested, and we conjecture that it will hold for all $\{m(a),\beta(a)\}$ models.
However, this might not be true for cases where $m(a)$ and/or $\beta(a)$ are very steep functions of $a$, leading to a very rapid evolution of the screening condition with density.
\item \textit{Choice of parameters.} Throughout this work, we fixed a number of parameters. Probably, the most influential ones are the halo cutoff radius $x_{\rm cut} = 10$ and defined $\mu_{200}$ as $\bar\gamma(\mu_{200})/\gamma{\rm max} = k$ with $k=1/2$. The latter choice is arguably quite natural. In spite of that, we do not expect our results to change dramatically if a (slightly) higher or lower value of $k$ is taken since the model curves are expected to shift equally. The choice of $x_{\rm cut}$ affects our results more drastically, as can be inferred from Fig.~\ref{fig:gamma_vs_R}. A lower value of $x_{\rm cut}$ leads, for instance, to a decrease in the extent of the partially screened region (i.e., an increase in $\alpha$). Nevertheless, since all models are affected by this, the precise choice of $x_{\rm cut}$ (within reason) does not alter our conclusions.
\item \textit{Practicality.} We focused merely on the force profiles directly, leaving the question how practical the study is since $\gamma(r)$ and $\bar\gamma$ are not directly observable. Although this is not wrong, we caution that the fifth force triggers many observable phenomena, as has been shown in various {\it N}-body simulations. In particular, $\gamma(r)$ can be related to the ratio of the dynamical mass to the lensing mass \citep{2010PhRvD..81j3002S,2011PhRvL.107g1303Z,2012ApJ...756..166W}
\begin{align}
\gamma(r) = \frac{M_{\rm Dyn}(<r)}{M_{\rm Lensing}(<r)} - 1,
\label{eq:gamma_masses}
\end{align}
where 
\begin{align}
M_{\rm Dyn} \equiv& 4 \pi \int r^2 \nabla^2 \left(\frac{\Phi+\Psi}{2}\right) \dd r\\
M_{\rm Lensing} \equiv& 4 \pi \int r^2 \nabla^2 \left(\frac{\Phi-\Psi}{2}\right) \dd r.
\end{align}
Here, $(\Phi+\Psi)/2$ is the dynamical potential, $(\Phi-\Psi)/2$ is the lensing potential, and $\Phi$ and $\Psi$ are the two metric potentials. For the models considered in this paper, the lensing potential is the same as in $\Lambda$CDM, but there are several models where the lensing potential is modified as in the Galileon, for example\footnote{While this paper was in the final stages of completion, \citet{2015arXiv150503468B} showed that for the Cubic Galileon model, the lensing potential within clusters is practically unmodified.}.
In practice, however, it is not possible to measure the potentials directly. Instead, the
dynamical mass is usually obtained via the cluster velocity dispersion through a power-law relation obtained from ($\Lambda$CDM) {\it N}-body simulations \citep[e.g.,][]{2007ApJ...669..905B}. The lensing mass can be read from the stacked density profiles assuming a constant mass-richness relation \citep{arXiv0709.1159J,2013SSRv..177...75H}. This means that our mass-weighted average of the gravitational enhancement $\bar\gamma$ is directly proportional to the ratio of the dynamical to the lensing mass as has been shown by \citet{2010PhRvD..81j3002S}\footnote{In fact, \citet{2010PhRvD..81j3002S} derive the relation $M_{\rm Dyn, observed} = (\bar\gamma + 1)^{3/5} M_{\rm Lensing}$ (see their Eq.~(64).).}.
\end{itemize} 
In conclusion, our work can be seen as a first step toward a truly unified description of SMGs on cluster scales. Naturally, this first step can be tuned further and extended to improve its precision and include more models.

\section{Conclusions}
\label{sec:conclusion}
Starting from the simple idea that screened modified gravity theories in general possess three regimes within nonlinear astrophysical structures, a fully screened, an unscreened, and a partially screened regime, we investigated whether one can specify, independently of the SMG model, at which halo mass range these regimes are located and how extended they are.

These two questions can be answered as follows:
\begin{itemize}
\item \textit{It is possible to formulate the mass of the partially screened regime as a function of various model parameters}. We demonstrated this empirically for the Symmetron, the Hu-Sawicki $f(R),$ and the nDGP models (\S\ref{sec:rescaling}). In addition, we found a semi-analytic derivation of this functional form in the $\{m(a),\,\beta(a)\}$-parametrization of \citet{2012PhRvD..86d4015B}. This allowed us to expand our method easily to the Dilaton model.
\item \textit{The extent of the partially screened regime is an intrinsic property of a particular screening mechanism.} We found that the mass range of the partially screened region differs significantly between the three example models but are (approximately) constant within a particular model.
\end{itemize}

In spite of the differences in the nature of their screening mechanisms (see \S~\ref{sec:smg-topography}), we characterize the Symmetron, the Hu-Sawicki $f(R),$ and the nDGP models with three parameters: \textit{(i)} the center of the partially screened regime $\mu_{200}$, \textit{(ii)} the maximal enhancement of the fifth force $\gamma_{\rm max}$, and the \textit{(iii)} effectiveness of the screening $\alpha$ (which controls the extent of the partially screened regime). We derived the first two parameters analytically in the $\{m(a),\,\beta(a)\}$-framework, which includes the Symmetron, the Hu-Sawicki $f(R)$, and the Dilaton models.

Quantifying observational constraints in the proposed $(\mu_{200},\,\gamma_{\rm max},\,\alpha)$-parametrization thus allows several screened modified gravity models to be constrained in bulk instead of having to consider them -- and their respective parameter space -- individually.

\begin{acknowledgements}
We thank the anonymous reviewer for useful comments.
MG thanks the Astrophysics department at the University of Oxford for
their hospitality.
DFM acknowledges support from the Research Council of Norway through grant $216756$. HAW is supported by the BIPAC and the Oxford Martin School. 
\end{acknowledgements}

\appendix

\section{NFW equations}
\label{sec:nfw-profile}
The Navarro-Frenk-White (NFW) profile is an empirically derived halo profile \citep{Navarro1995,Navarro1996,Navarro1997} from {\it N}-body simulations.
The NFW density profile reads as a function of normalized distance from the center $x \equiv r / R_{\rm vir}$:
\begin{equation}
\frac{\rho(x)}{\rho_{c0}} = \frac{\Delta_{\rm vir} c^2 g_{\rm NFW}(c)}{3 x (1 + c x)^2}
\label{eq:nfw_rhox}
\end{equation}
where
\begin{equation}
g_{\rm NFW}(c) = \frac{1}{\ln(1+c) - c / (1 + c)}.
\label{eq:gnfw}
\end{equation}
Here, $c$ is the concentration parameter,  which we link to $M_{200}$ through the observed relationship found in  \cite{2001MNRAS.321..559B}:
\begin{equation}
c = 9 \left(\frac{M_{200}}{3.2 \times 10^{12}\Msh}\right)^{-0.13}\;.
\label{eq:NFW_c-M_relation}
\end{equation}
Throughout this work we define $\Delta_{\rm vir}\equiv 200$, which means $R_{\rm vir}=R_{200}$ and $M_{\rm vir} = M_{200} = 200 \times 4/3 \pi R_{\rm vir}^3 \rho_{c0}$.

Although the density of the NFW profile is diverging for $x\rightarrow 0$, the enclosed mass within a certain radius does not \citep{Lokas2001}. This leaves the mass fraction within $[x\pm \dd x/2]$ as
\begin{equation}
\frac{\dd M}{M(<x_{\rm cut})} = \frac{c^2 x g_{\rm NFW}(c x_{\rm cut})}{(1 + c x)^2}\dd x\;,
\label{eq:dMnfw}
\end{equation}
where $x_{\rm cut}$ is defined below Eq.~\eqref{eq:gammabar_def}. This will be useful when considering mass-weighted quantities. 

To calculate the screening condition we also need the expression for $\Phi_N$. For a NFW halo, we find
\begin{equation}
|\Phi_N|(x) = \frac{\Omega_m}{a}\frac{(H_0R_{\rm vir})^2\Delta_{\rm vir} g_{\rm NFW}(c)}{x}\log(1+cx)\;.
\end{equation}

\section{DGP solutions for NFW halos}
\label{sec:dgp_nfw}
In this appendix we present the equations for $\gamma$ for the DGP model.

For a NFW density profile (Eq.~\eqref{eq:nfw_rhox}) halo, we have
\begin{align}
\frac{\int_0^r \delta_m(r) r^2 dr}{r^3}  = \frac{\Delta_{\rm vir}}{3x^3}\left(\frac{\log(1+cx)  - \frac{cx}{1+cx}}{\log(1+c) - \frac{c}{1+c}}\right) = \frac{\Delta_{\rm vir}g_{\rm NFW}(c)}{x^3g_{\rm NFW}(cx)}
\end{align}
where $x = \frac{r}{R_{\rm vir}}$. This leads to the fifth force profile being given by
\begin{align}
\gamma &= \gamma_{\rm max}\left[\frac{2(\sqrt{1+\epsilon(x)}-1)}{\epsilon(x)}\right]
\label{eq:ngdp_nfw_gamma}
\end{align}
where $\gamma_{\rm max} = \frac{1}{3\beta_{\rm DGP}}$ and
\begin{align}
\epsilon(x) &= \frac{8(r_cH_0)^2\Omega_m}{9\beta_{\rm DGP}^2}\frac{\Delta_{\rm vir}g_{\rm NFW}(c)}{x^3g_{\rm NFW}(cx)}.
\end{align}

\bibliography{references}

\end{document}